\documentstyle[preprint,aps]{revtex}

\begin{document}

\title{Time-delayed Spatial Patterns in a Two-dimensional Array of Coupled Oscillators} 

\author{Seong-Ok Jeong, Tae-Wook Ko\cite{TWKo_mail}, and Hie-Tae Moon}
\address{Department of Physics, Korea Advanced Institute 
	of Science and Technology, Taejon 305-701, Korea}
\date{\today}
\maketitle
\begin{abstract}
We investigated the effect of time delays on phase configurations 
in a set of two-dimensional coupled phase oscillators. Each oscillator is allowed to 
interact with its neighbors located within a finite radius, which serves as a 
control parameter in this study. It is found that distance-dependent time-delays induce various patterns including traveling rolls, square-like and rhombus-like patterns, spirals, and targets. We analyzed the stability boundaries of the emerging patterns and briefly pointed out the possible empirical implications of such time-delayed patterns.
\\
PACS numbers: 47.54.+r, 05.45.Ra, 82.40.Ck, 89.75.Kd \\
\pacs{}
\end{abstract}

\maketitle
\narrowtext


Spatiotemporal patterns arise in numerous physical, chemical, and biological systems \cite{cross}. 
The brain, one of the most complex systems, is now also known to generate spatiotemporal patterns such as plane waves and spirals \cite{cat,haken,prechtl,hallu,ermen}. Early studies, in this context, have concentrated on exploring the dynamics arising in a set of coupled oscillators\cite{ermen,kura,tass,somp,shp2,swan}. It has been, however, generally assumed that interactions among the individual oscillators are instantaneous. Recently, noting that an inclusion of time delay is more natural in realistic systems, 
several authors have investigated the effects of time delay, and have found the consequent dynamic phenomena such as multistability, desynchronization, clustering, amplitude death, anticipated synchronization and slow switching \cite{somp,shp2,swan,schuster,strogatz,skim,reddy,other_rsch}. It has been also found recently that a time delay, proportional to a distance between elements, could produce a propagating structure in a one-dimensional coupled-oscillators\cite{pre}.  

In this paper, we discuss further the effect of time delays on the dynamics in a complex system in two dimensions, employing a two-dimensional coupled oscillator model\cite{cat,haken,prechtl,hallu,ermen}. It is found that time delays among constitutive elements alone can induce definite spatiotemporal patterns in two-dimensional complex system. 

We start with the following coupled oscillator model equations for phase variables $\theta_{ij}(t)$ ($i$ and $j$ are integers):
\begin{equation}
\dot{\theta}_{ij}(t) = w + \frac{K}{N(r_{0})}
\sum\limits_{k,l} ^{0 <r_{kl,ij} \le r_{0}} W(r_{kl,ij}) 
\sin \left [\theta_{kl}(t-\frac{r_{kl,ij}}{v})
- \theta_{ij}(t) \right],  
\label{oreq}
\end{equation}
with $N(r_{0})=\sum\limits_{k,l} ^{0<r_{kl,ij} \le r_{0}} W(r_{kl,ij})$.
Here $r_{kl,ij}$ denotes the distance between element $(i,j)$ and element $(k,l)$.
The metric properties of the ensemble are fixed by setting $r_{kl,ij}=\sqrt{(i-k)^2+(j-l)^2}$.
In this model an interaction is non-local and time delayed, characterized by coupling length $r_{0}$ and signal propagation speed $v$, and time delays are expressed by $r_{kl,ij}/v$. Assuming that an individual element interacts with its neighbors located within a finite radius, we introduce a coupling radius denoted by $r_{0}$; We also discussed the case of $r_{0} \rightarrow \infty$ in this study. Further we introduce a weighting function $W(r_{kl,ij})$ which mimics the interaction strength depending on a distance between interacting elements. Since there is no law about $W(r_{kl,ij})$, as far as we know, we choose $W(r_{kl,ij})=1/r_{kl,ij}$ assuming that the Gauss' law of interactions in two dimensional cases is also valid here\cite{Gauss}.

The above equation without time delays is a model of coupled limit cycle oscillators 
in the so-called phase approximation \cite{kura}. 
It is known that, in the absence of delays $r_{kl,ij}/v$ , 
for any positive values of the coupling coefficients $K$, 
all the oscillators finally form a planar solution where all the oscillators are synchronized, oscillating with frequency equal to that of the individual oscillators \cite{kura}.  
The present study of the model (\ref{oreq}), on the other hand, shows that the final steady states correspond to planar solutions 
with frequencies different from $w$ or to frequency-synchronized nontrivial phase configurations.     

To find out possible phase configurations, 
let us assume steady solutions in which all the oscillators have the same frequency $\Omega$.
Then the solutions of the model can be written as $ \theta_{ij}(t) = \Omega t + \phi_{ij}$.
Substitution of these solutions into Eq. (\ref{oreq}) yields 
\begin{equation}
\Omega=w+\frac{K}{N(r_{0})} \sum\limits_{k,l} ^{0<r_{kl,ij} \le r_{0}} 
\frac{1}{r_{kl,ij}}
\sin(\phi_{kl}-\phi_{ij}-\Omega r_{kl,ij}/v).
\label{eq1}
\end{equation}
This self-consistency condition requires that the sums
$\frac{1}{N(r_{0})}\sum\limits_{k,l} ^{0<r_{kl,ij} \le r_{0}} \frac{1}{r_{kl,ij }}
\sin(\phi_{kl}-\phi_{ij}-\Omega r_{kl,ij}/v)$ be same for all $(i,j)$ \cite{pre}.
Let us consider the situation in the continuum limit. 
Differentiating the term with respect to $\vec{r}[=(i,j)]$, we get the following.

\begin{equation}
\frac {\partial \phi(\vec{r'})} {\partial \vec{r'}} = \frac {\partial \phi(\vec{r})} {\partial 
\vec{r}} \textrm{ , for all } 
\vec{r} \textrm{ and } \vec{r'}.
\end{equation}
This implies that, in the continuum limit, 
$\phi(\vec{r}) = \gamma\vec{r} \cdot \vec{e}+\phi_{0}$ is a unique solution,
up to the arbitrariness of $\phi_{0}$ and $\vec{e}$, as long as
$\vec{e}$ is a unit vector embedded in the lattice plane.
The case of $\gamma=0$ corresponds to synchronized planar solutions, while nonzero
$\gamma$'s make roll solutions traveling in the direction of $\vec{e}$ 
with the phase velocity $\Omega/\gamma$.

However, the space discreteness of the system admits topological defects, so other phase configurations are also possible \cite{rrr}.
In this study, we set $w=\pi/5$, and 
solved the equation on a 64 by 64 square plane 
($1 \le i \le 64$, $1 \le j \le 64$) with periodic boundary conditions.
We checked that different size of lattice does not alter the result.
We tried several kinds of initial conditions. 
As the initial conditions in phase of the form $\theta_{ij}(0) = \theta_0
+ \eta_{ij}$ in this study, where $\eta_{ij}$ denotes a random number, we
considered two different cases: (i) $ 0 \le \eta_{ij} <2 \pi $, 
and (ii) $0 \le \eta_{ij} < 10^{-6}$.
 In addition, for the dynamic
history of individual oscillators for $t < 0$, we considered two situations
(i) where the oscillators were stationary, and (ii) where the oscillators have evolved independently from one another with their natural frequency $w$.  

Figure 1, which is made with the parameters $K=0.6$ and $r_{0}=10$, 
shows effects of the delay $1/v$.
When the delay $1/v$ is small ($< 0.6$),
only planar solutions($\gamma=0$) occur. 
When the delay is large enough($>1.1$), planar solution is not available but patterns 
with nonzero $\gamma$'s occur.
In the intermediate region ($0.6 \le 1/v \le 1.1$), planar solutions as well as patterns are possible. 
The lines in Fig.1 are the numerically obtained solutions of Eq. (\ref{eq1}) with 
$\phi_{ij}=\gamma i$, $\gamma = 0, \frac{5\pi}{32}, \frac{6\pi}{32}$, and $ \frac{7\pi}{32}$.
We see that the wavelengths of patterns become shorter generally as delay increases.

Figure 2 provides the phase diagrams of the model. 
If the coupling strength $K$ is above $K_{c2}$ (denoted by open squares), 
only a synchronized planar solution appears, but
if $K$ is below the critical value $K_{c1}$ (denoted by filled squares), planar solution is not available and
the system evolves into patterns such as rolls [as in Fig. 3(a)], or other phase
distributions [as in Figs. 3(b) - 3(d)] depending on initial configurations.
In the regions of $K_{c1} < K < K_{c2}$, we see both patterns and planar solutions depending on initial configurations. 
Figure 4 shows the patterns that appear only in the limited regions in the phase diagram.
When $r_{0}$ is relatively small ($\lesssim 5$), in other words,
when the effect of space discreteness becomes important,
spirals occur more frequently [See Fig. 4(a)] than other configurations.  
Furthermore, the model sometimes organizes target patterns from slightly perturbed homogeneous phase configurations for parameter $K \sim K_{c1}$ [see Fig. 4(b)]. 
With $K \sim K_{c2}$, there emerge patterns embedded in a planar solution as shown in Fig. 4(c).

Note that the arrangement of the phase singularities plays an essential role
in forming the patterns of Figs. 3(b) - (d). The singularities are possible due to the discreteness of 
the space as in spiral patterns, but are dynamically arranged in a rhombus [Fig. 3(b)] or a 
square lattice [Fig. 3(c)]; the phases of oscillators make rotating waves around those singularities \cite{movie}. However, rolls are more prevalent than squares or rhombus patterns in the entire region of the diagram. 

There seem to exist three kinds of multistabilities in our model. First, as stated above,
we can see planar solutions, rolls or other phase distributions 
for the same parameters depending on initial configurations [see Figs. 1-3].
Second, there are multiple states with different wavelengths 
and frequencies even for the same parameters in some cases, 
which is why we see multiple $\Omega$'s of patterns in Fig. 1.
This kind of multistability is also reported in one-dimensional lattices \cite{pre}.
Third, different initial configurations make the different angles between two
primitive translation vectors of patterns like Figs. 3(b) and 3(c).
It seems that multistabilities are the major characteristics of time delayed systems 
\cite{schuster,strogatz,skim}.

In the systems of two-coupled oscillators \cite{somp,schuster}, or globally coupled oscillators with uniform time delays \cite{strogatz}, the dynamics are determined by
the value of $\Omega\tau$, where $\Omega$ is the frequency of the synchronized state and $\tau$ is a time delay between oscillators.
The physical meaning of the value is clear. It denotes the virtual phase difference between oscillators due to delay.
If it is near $2n\pi$, where $n$ is an integer, each oscillator can keep pace with others to make the whole system synchronized, otherwise it cannot \cite{strogatz}.
Likewise, we suggest that the dynamics of our model is determined by the 
quantity $\Omega r_{0} /v$, which we call $\Theta$ here, 
where $\Omega$ is a frequency of a synchronized planar solution.
Our conjecture is that there exists some value $\Theta_{c}$ so
that planar solutions are possible only for $\Theta < \Theta_{c}$ \cite{cmtt}. 

We measured $\Theta_{c}$ directly from numerical simulations 
using Eqs. (\ref{oreq}) and (\ref{eq1}) in Fig. 5.
We see that $\Theta_{c}$ is a nearly constant value ($\sim 2.26$), 
from which we can calculate $K_{c1}$ in Fig. 2.
Substituting the roll solution $\phi = \gamma x$ to 
the continuum limit of Eq.(\ref{eq1}) makes the following equation.
\begin{equation}
\Omega = w - \frac{K}{r_{0}} \int_{0}^{r_{0}} \sin (r\Omega /v) J_{0}(\gamma r) dr,
\end{equation}
where $J_{0}$ is a Bessel function.
In the case of planar solutions ($\gamma=0$), the above equation can be integrated to give
\begin{equation}
K = \frac{\Theta w}{2 \sin ^{2} \frac{\Theta}{2}} (1- \frac{\Theta}{wr_{0}\frac{1}{v}}).
\label{phdi}
\end{equation}
The solid lines in Figs. 2 (a) and (b) are obtained by substituting $\Theta=2.26$ to Eq.(\ref{phdi}),
and show good agreements with the simulation results.

It is useful to look at what the existence of constant $\Theta_{c}$ implies. 
It means the following. 
Coupled oscillators tend to be synchronized, 
but when too large delays make the $\Theta$ value go beyond the tolerance limit $\Theta_{c}$, 
they fail to become a synchronized planar solution and take other configurations. 
The occurrence of incoherence in the coupled oscillators 
with uniform time delay \cite{strogatz,skim} can now be explained in the same manner.

It is also noteworthy that, in the limit of $r_{0} \rightarrow \infty$, Eq. (5) converts to  
$K_{c} = \frac{\Theta_{c} w}{2 \sin ^{2} \frac{\Theta_{c}}{2}}$, which is independent of $1/v$.
This is surprising, for it means that we must consider the effects of time delay however small 
when the interaction length is infinite as in most physical systems. 

To summarize, we have investigated the effects of time delayed interactions in an ensemble
of coupled oscillators in two dimensions, and found that distance-dependent time delay alone can induce various
spatial patterns, while without delay or with uniform delay alone only synchronized planar solutions might appear. We have analyzed also that the stability of planar solutions can be
determined by the quantity $\Omega r_{0} /v$. 
Our study may find its relevance in recent experimental observations that non-locally interacting oscillators, which are especially 
probable for neurobiological systems, exhibit traveling rolls and spirals \cite{prechtl,ermen}. 
Finally, our results seem to suggest that time delay may play a significant role in the study of memory storage 
and information processing based on the spatiotemporal  
activities of neurons \cite{cat,haken,prechtl,hallu,ermen}. 

We thank  Hwa-Kyun Park for useful discussions.
This work was supported in part by the Interdisciplinary Research Program
(Grant No. R01-1999-00019) of the Korea Science and Engineering Foundation, and in part by 
the Brain Korea 21 Project in 2000. We also appreciate the support of the Korea 
Research Foundation (Grant No. KRF-2000-015-DP0097).

\pagebreak
\newpage
{\large \bf Figure Captions}
\begin{description}
\item[Fig. 1.] Synchronization frequency $\Omega$ as a function of a delay ($K=0.6, r_{0}=10$); 
squares indicate planar solutions; circles indicate patterns.
Lines are the solutions of Eq.(\ref{eq1}). See the text for details.  
\item[Fig. 2.] The phase diagrams of the model with $r_0 = 10$ in (a), and with $1/v=1$ in (b).
Above open squares, there is no pattern available, and only planar solutions appear. 
Between open and filled squares, the system evolves
into planar solutions or patterns according to initial conditions.
Below the filled squares, planar solution is not possible here, and only patterns appear.  
Solid lines are from Eq.(\ref{phdi}). See the text for details.
\item[Fig. 3.] Typical patterns generated in the model :
(a) a traveling roll, (b) a rhombus-like pattern, (c) a square-like pattern,
(d) coexistence of a rhombus-like pattern and a roll.       
$K=1.25$, $r_{0}=25$, and $1/v=1$,
for (a), (b), and (c). Note that planar solutions are also found for this set of parameter values.; $K=0.6$, $r_{0}=10$, and $1/v=1.1$, for (d).
See Fig. 4 (d) for the color coding.
\item[Fig. 4.]  Patterns obtained in simulations of the model :
(a) spirals ($K=1.0$, $r_{0}=4$, and $1/v=1$),
(b) a target pattern ($K=0.4$, $r_{0}=7$, and $1/v=1$),
(c) a spiral embedded in a planar solution ($K=0.8$, $r_{0}=8$, and $1/v=1$). In (c), the planar solution oscillates much more slowly than the spiral.
Figure (d) shows the color code used in this paper.
\item[Fig. 5.] The quantity $\Omega r_{0} /v$ determines the stability of planar solutions.
We cannot see planar solutions above the filled squares.
\end{description}

\begin{thebibliography}{99}
\bibitem{Gauss} We have also tried $W=1$ in this study, yielding the same qualitative results for pattern formation except that transition points shift slightly in phase diagrams.
\bibitem{cmtt} Note that the translational symmetry $\Theta \to \Theta + 2n\pi$ does not hold,
for $\Theta$ is related only with the maximum delay.
\end{thebibliography}
\end{document}